\documentclass[amsmath, amssymb, twocolumn, aps, prx, superscriptaddress, footinbib]{revtex4-1}
\usepackage{mathtools}
\usepackage{mathrsfs, amsmath, wasysym}
\usepackage{bbold}
\usepackage[mathscr]{eucal}
\usepackage{xfrac}
\usepackage{graphicx}
\usepackage{dcolumn}
\usepackage{bm}
\usepackage{color}
\usepackage{tabularx}
\usepackage{natbib}
\usepackage[colorlinks,linkcolor=blue,citecolor=blue,urlcolor=blue]{hyperref}
\usepackage[normalem]{ulem}
\usepackage[all]{hypcap}
\usepackage[dvipsnames,usenames,table]{xcolor}
\usepackage{braket,mleftright}

\newcommand{\nocontentsline}[3]{}
\newcommand{\tocless}[2]{\bgroup\let\addcontentsline=\nocontentsline#1{#2}\egroup}

\newcommand{\bk}{{\bf k}}

\newcommand{\bq}{{\bf q}}

\newcommand{\bz}{{\bf z}}

\newcommand{\bb}{{\bf b}}

\newcommand{\bj}{{\bf j}}

\newcommand{\bn}{{\bf n}}

\newcommand{\bM}{{\bf M}}
\newcommand{\bN}{{\bf N}}

\newcommand{\bh}{{\bf h}}

\newcommand{\btau}{\boldsymbol{\tau} }
\newcommand{\bsigma}{\boldsymbol{\sigma} }

\newcommand{\be}{\begin{equation}}
\newcommand{\ee}{\end{equation}}
\newcommand{\beg}{\begin{gather}}
\newcommand{\eeg}{\end{gather}}
\newcommand{\beq}{\begin{eqnarray}}
\newcommand{\eeq}{\end{eqnarray}}
\newcommand{\bea}{\begin{align}}
\newcommand{\eea}{\end{align}}
\newcommand{\beqq}{\begin{eqnarray*}}
\newcommand{\eeqq}{\end{eqnarray*}}

\newcommand{\up}{\uparrow}
\newcommand{\down}{\downarrow}

\begin{document}

\title{Layer dipole magnetoelectric polarizability of antiferromagnetic bilayers}

\author{H. Radhakrishnan}
\affiliation{Department of Physics, Drexel University, Philadelphia, PA 19104, USA}%
\affiliation{Department of Materials Science \& Engineering, Drexel University, Philadelphia, PA 19104, USA \looseness=-1}%

\author{C. Ortix}
\affiliation{Dipartimento di Fisica ``E. R. Caianiello'', Universit\`a di Salerno, IT-84084 Fisciano (SA), Italy}%
\affiliation{CNR-SPIN, I-84084 Fisciano (Salerno), Italy, c/o Universit\'a di Salerno, I-84084 Fisciano (Salerno), Italy}

\author{J. W. F. Venderbos}
\affiliation{Department of Physics, Drexel University, Philadelphia, PA 19104, USA}%
\affiliation{Department of Materials Science \& Engineering, Drexel University, Philadelphia, PA 19104, USA \looseness=-1}%

\begin{abstract}
In this paper we study magnetoelectric effects in two-dimensional magnetic bilayers and introduce the notion of a layer dipole magnetoelectric polarizability. This magnetoelectric polarizability describes the magnetization response to an applied electric field perpendicular to the bilayer. As such, it represents the electric analog of the spin magnetoelectric polarizability, governing the charge polarization response to an applied Zeeman field. Starting from the orbital magnetization produced by a perpendicular displacement field, we derive a microscopic expression for the layer dipole magnetoelectric polarizability and apply it to two minimal models for bilayer magnets, i.e., a buckled square lattice model and a magnetic topological insulator model. In the case of the buckled square lattice model we show that the layer dipole magnetoelectric polarizability has a (quasi-)topological contribution, revealing a topological magnetoelectric response of two-dimensional antiferromagnets associated with the layer pseudospin degree of freedom. 
\end{abstract}

\date{\today}

\maketitle

{\it Introduction.}---The study of two-dimensional (2D) materials is a subject of great current interest. Prominent examples of materials at the center of this frontier are graphene multilayers~\cite{CastroNeto:2009p109,Balents:2020p725,Andrei:2020p1265,Andrei:2021p201,Mak:2022p686,Pantaleon:2023p304,Bernevig:2024p38,Nuckolls:2024p460}, transition-metal dichalcogenides~\cite{Li:arXiv2025}, as well as a variety of layered van der Waals magnets~\cite{Burch:2018p47,Gibertini:2019p408,Xie:2022p9525}. The van der Waals magnets comprise a relatively recent addition to the family of 2D crystals and have seen a surge of activity following the observation of magnetic order in few-layer and single-layer samples~\cite{Huang:2017p270,Gong:2017p265}. The interest in (quasi-)2D materials originates from the rich landscape of phenomena they exhibit, ranging from electronic band topology~\cite{Qian:2014p1344,Po:2018p031089,Song:2019p036401,Wu:2019p086402,Pan:2020p033087,Devakul:2021p6730} to unusual quantum states of matter with strong electron correlations (e.g. superconductivity~\cite{Cao:2018p43,Zhou:2021p434,Zhou:2022p774,Zhang:2022p268,Xia:2025p833,Guo:2025p839}, magnetism~\cite{Gong:2017p265,Huang:2017p270}, and quantum anomalous Hall states~\cite{Chang:2013p167,Deng:2020p895,Li:2019peaaw5685,Zhao:2020p419,Sharpe:2019p605,Serlin:2020p900,Chen:2020p56,Polshyn:2020p66,Li:2021p641,Park:2023p74,Xu:2023p031037,Lu:2024p759}). Further appeal derives from the availability of versatile heterostructure engineering techniques, as well as the ability to control key electronic properties such as electron density and layer bias by electrostatic gating. 

In 2D multilayer materials the layer degree of freedom represents a pseudospin akin to valley pseudospin or physical spin, and is increasingly recognized as a powerful source of additional tunability and control~\cite{Pesin:2012p409,Xu:2014p343,Gao:2020p077401,Zheng:2024p8017,Fan:2024p7997,Hu:2024pL201403,Hu:2025arXiv}. Since the layer pseudospin has the symmetry of an electric dipole moment, it couples to electric fields and can thus be exploited for generating electric or magnetic responses specific to 2D multilayers. In the case of van der Waals magnets, the layer degree of freedom provides a compelling basis for exploring magnetism in reduced dimensions, in particular the realization of new types of field-controlled magnetic behavior.  One area of great importance for magnetic information storage devices is the cross-coupling of magnetic and electric response properties, i.e., the electric response of a material to applied magnetic fields or vice versa~\cite{Fiebig:2005pR123,Spaldin:2005p391}. Indeed, the ability to electrically control and switch magnetic behavior is one of primary aims of spintronics and motivates the ongoing search for materials showing strong magnetoelectric coupling~\cite{Tokura:2014p076501,Fiebig:2016p16046,Fiebig:2016p16046,Baltz:2018p015005,Manchon:2019p035004}. In light of this search it is natural to ask whether magnetism in 2D multilayers can lead to magnetoelectric effects that rely on the layer pseudospin, and are thus specific to two dimensions~\cite{Gong:2013p2053,He:2020p1650,Tao:2024p096803,Fan:2024p7997,Hu:2025arXiv}.

\begin{figure}
	\includegraphics[width=\columnwidth]{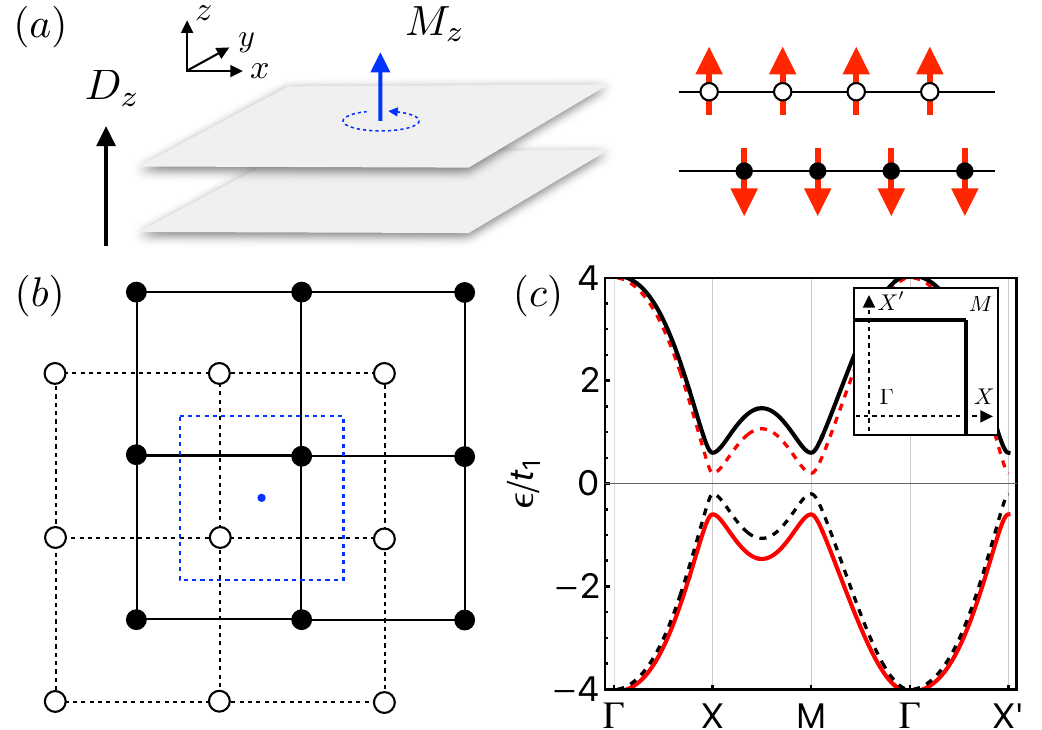}
	\caption{(a) Illustration of the layer dipole magnetoeletric effect in bilayer antiferromagnets. An orbital magnetization ($M_z$) develops in response to a perpendicular displacement field $D_z$. The 2D antiferromagnet is formed from two oppositely aligned easy-axis ferromagnetic layers (shown on the right). (b) The buckled square lattice with two sublattices (white and black sites) displaced in the negative and positive $z$ direction (i.e., out-of-plane direction), respectively. The inversion center at the origin of the unit cell is marked as a blue dot. (c) Energy spectrum of the buckled square lattice model in the presence of displacement field, as calculated from Eqs.~\eqref{eq:H_k} and~\eqref{eq:buckled-square}. Here we have set $\lambda=0.6W$, $N_z=0.4W$, and $edD_z=0.2W$, with $W\equiv  \hbar^2/2m_ea^2$ (we always set $t_1 =W$). The inset shows the Brillouin zone. }
	\label{fig1}
\end{figure}

In this paper we address this question by studying linear magnetoelectric effects in bilayer antiferromagnets with ${\mathcal P}{\mathcal T}$ symmetry~\cite{Smejkal:2017p106402,Wang:2017p115138,Godinho:2018p4686,Du:2020p022025,Wang:2021p277201,Liu:2021p277202,Wang:2023p056401,Tao:2024p096803,Liang:2024p256901}. We demonstrate in particular that bilayer antiferromagnetic insulators can develop an out-of-plane orbital magnetization ($M_z$) in response to a perpendicular electric displacement field ($D_z$), as shown schematically in Fig.~\ref{fig1}(a). Since the coupling of the displacement field to the layer pseudospin resembles the Zeeman coupling between magnetic field and spin, the associated magnetoelectric polarizability, defined as $\alpha_{zz }  = \partial M_z / \partial D_z$ and dubbed layer dipole magnetoelectric polarizability, can be viewed as the dual of the spin magnetoelectric polarizability~\cite{Gao:2018p134423}, which describes the linear charge polarization response to a Zeeman field. 

Below we introduce a general class of microscopic models to describe bilayer magnets and derive a simple expression for the magnetoelectric polarizability $\alpha_{zz}$. The general theory of magnetoelectric response is then applied to two specific models, all motivated by known material realizations. These include a thin film magnetic topological insulator model and a buckled square lattice model applicable to the antiferromagnet CuMnAs~\cite{Smejkal:2017p106402,Wang:2017p115138,Godinho:2018p4686}. The analysis of these specific models reveals that $\alpha_{zz}$ can have a (quasi)topological contribution when the energy bands of the nonmagnetic bilayer feature 2D Dirac points.

{\it Orbital magnetization of bilayers: general theory.}---To construct a minimal microscopic model for a bilayer antiferromagnet, we consider electrons with layer ($\alpha=1,2$) and spin ($\sigma=\up,\down$) degrees of freedom and introduce the corresponding electron destruction operators $c_{\bk \alpha \sigma}$. In this basis the Hamiltonian is written as $H = \sum_\bk  c^{\dagger}_\bk H_\bk c_\bk$ with $H_\bk$ given by the general form
\be
H_\bk = f_{x,\bk} \tau^x + f_{y,\bk} \tau^y + \tau^z\bn_{\bk} \cdot \bsigma + \Delta \tau^z.  \label{eq:H_k}
\ee
Here $\btau =(\tau^x,\tau^y,\tau^z)$ and $\bsigma =(\sigma^x,\sigma^y,\sigma^z)$ are two sets of Pauli matrices corresponding to layer ($\tau^z=\pm1$) and spin ($\sigma^z=\pm1$), respectively, and $\Delta = ed D_z$ is the displacement field multiplied by the electron charge $e$ and the distance between the layers $d$. The coefficients $f_{x/y,\bk}$ and $\bn_\bk = (n_x,n_y,n_z)$ are real momentum-dependent functions and parametrize a broad class of bilayer antiferromagnets. Below we examine three specific examples. The final term in Eq.~\eqref{eq:H_k} describes the coupling of the displacement field $D_z$ to the layer dipole $ed\tau^z$. 

In its nonmagnetic state the bilayer has an inversion symmetry ${\mathcal P}$, which exchanges the layers and hence requires that the Hamiltonian satisfies $\tau^x H_\bk \tau^x = H_{-\bk }$. Inversion symmetry therefore implies $(f_{x,-\bk},f_{y,-\bk})=(f_{x,\bk},-f_{y,\bk})$ and $\bn_{-\bk}=-\bn_{\bk}$. The displacement field $D_z$ clearly breaks inversion symmetry. Time-reversal symmetry (${\mathcal T}$) imposes the constraint $\sigma^y H^*_\bk \sigma^y = H_{-\bk }$ and leads to the same conditions on $f_{x/y,\bk}$ and $\bn_\bk $. It is then straightforward to show that with $D_z$ set to zero (i.e., no displacement field) $H_\bk$ in Eq.~\eqref{eq:H_k} describes the most general bilayer Hamiltonian compatible with ${\mathcal P}{\mathcal T}$ symmetry, i.e., the product of inversion and time-reversal. 

The antiferromagnetic (AFM) state of the bilayer is formed from two ferromagnetically ordered layers with anti-aligned moments, and is described by a N\'eel vector $\bN $. This is schematically shown in Fig.~\ref{fig1}(a). For the purpose of this work, we assume that the N\'eel vector points in the $z$ direction, i.e., $\bN = N_z \hat\bz$, such that it enters the Hamiltonian $H_\bk$ of Eq.~\eqref{eq:H_k} as part of $n_z$. The AFM N\'eel state breaks ${\mathcal P}$ and ${\mathcal T}$ symmetry, but preserves the product symmetry ${\mathcal P}{\mathcal T}$. As a result, a linear magnetoelectric coupling between the displacement field $D_z$ and the magnetization $M_z$ of the form $M_z = \alpha_{zz}D_z$ is allowed, such that the magnetoelectric polarizability $\alpha_{zz}$ is a generic property of the AFM bilayer. 

We now turn to a microscopic calculation of the polarizability $\alpha_{zz }$. To obtain an expression for $\alpha_{zz }$, we compute the orbital magnetization $M_z$ for the class of models defined by Eq.~\eqref{eq:H_k} and take the derivative with respect to $D_z$. The orbital magnetization is given by the general formula~\cite{Xiao:2005p137204,Thonhauser:2005p137205,Ceresoli:2006p024408}
\be
M_z = \frac{e}{2\hbar} \epsilon_{ij} \int \frac{d^2 \bk}{(2\pi)^2} \sum_n \text{Im} \langle \partial_i u_{\bk n} | H_\bk + \varepsilon_{\bk n} | \partial_j u_{\bk n} \rangle ,\label{eq:M_z-def}
\ee
where $\varepsilon_{\bk n}$ are the energies with band index $n$, and $| u_{\bk n} \rangle $ are the corresponding eigenstates. (We suppress momentum dependence in what follows.) The sum on $n$ is over all occupied bands and $\partial_i \equiv \partial / \partial k_i $, with $i,j=x,y$. When applied to $H_\bk$ of Eq.~\eqref{eq:H_k}, the integrand on the r.h.s. of Eq.~\eqref{eq:M_z-def} can be recast in terms of the band energies $\varepsilon_{ n}$ and the Hamiltonian itself, in particular the vector $\bn_\bk$ (see Appendix~\ref{app:Mz-4-band} for details). The four energy bands of $H_\bk$ are given by $(\varepsilon_{1},\varepsilon_{2},\varepsilon_{3},\varepsilon_{4})=(-\varepsilon_{+},-\varepsilon_{-},\varepsilon_{-},\varepsilon_{+})$, where $\varepsilon_{\pm}$ is defined as $\varepsilon_\pm \equiv \sqrt{f^2 +(|\bn|\pm \Delta)^2}$ and $f^2 \equiv f^2_x + f^2_y $. Note that this ordering of energy bands assumes $\Delta>0$; when $\Delta<0$ one simply exchanges $\varepsilon_{+} \leftrightarrow \varepsilon_{-}$. As expected, in the absence of a displacement field ($D_z=0$) the energy spectrum has a manifest twofold degeneracy at each $\bk$, even in the presence of AFM N\'eel order, as a consequence of ${\mathcal P}{\mathcal T}$ symmetry. As detailed in Appendix~\ref{app:Mz-4-band}, inserting the energies and eigenstates of Eq.~\eqref{eq:H_k} into Eq.~\eqref{eq:M_z-def} then yields 
\be
M_z = \frac{e^2dD_z }{\hbar}  \int \frac{d^2 \bk}{(2\pi)^2}  \frac{\bn \cdot \partial_ x \bn \times \partial_y \bn}{(\varepsilon_{+} +\varepsilon_{ -} )^3}  \Gamma_{\varepsilon_{\pm},\Delta}, \label{eq:M_z}
\ee
with $\Gamma_{\varepsilon_{\pm},\Delta} $ given by
\be
\Gamma_{\varepsilon_{\pm},\Delta} = \frac{ 4 \Delta^2-(\varepsilon_++ \varepsilon_-)^2 }{ \varepsilon_+ \varepsilon_-}. \label{eq:Gamma}
\ee
Equation~\eqref{eq:M_z} gives the orbital magnetization as a function of $D_z$ for the general class of models described by Eq.~\eqref{eq:H_k}. Note that it directly follows from~\eqref{eq:M_z} that $M_z$ vanishes when $D_z=0 $, as required by ${\mathcal P}{\mathcal T}$ symmetry.
To obtain the polarizability $\alpha_{zz } $, we take the derivative of $M_z$ with respect to $D_z$ and then set $D_z=0$, which yields the result
\be
\alpha_{zz }  =  -\frac{e^2d}{\hbar}  \int \frac{d^2 \bk}{(2\pi)^2}  \frac{\bn \cdot \partial_ x \bn \times \partial_y \bn}{(f^2 +\bn^2)^{3/2} }. \label{eq:alpha_zz}
\ee
This expression for the magnetoelectric polarizability, which describes the orbital magnetization response to an applied displacement field, and relies on the coupling of the latter to the layer dipole pseudospin, is the first key result of our work. The  integrand of Eq.~\eqref{eq:alpha_zz}, in particular the numerator, signals that $\alpha_{zz } $ is sensitive to quantum geometric properties of the energy band structure of the bilayer antiferromagnetic. The specific models we now examine confirm this, and expose the (quasi)topological contribution to the magnetoelectric response captured by $\alpha_{zz } $.

{\it Application to buckled square lattice model.}---As a first example of a bilayer antiferromagnet, we consider the bilayer (or buckled) square lattice model shown in Fig.~\ref{fig1}(b). This bilayer model was considered in Ref.~\onlinecite{Young:2015p126803} as a minimal model for a (spin-orbit coupled) Dirac semimetal in 2D. The space group of the nonmagnetic buckled square lattice is $P4/nmm$ and the effect of buckling on the electronic structure is to allow for a spin-orbit term in the Hamiltonian given by $2\lambda \tau^z (\sin k_y \sigma^x-\sin k_x \sigma^y)$. The full Hamiltonian, including N\'eel order along the $\hat z$ direction, is of the form \eqref{eq:H_k}, with $f_{x,\bk} $ and $ \bn_{\bk}$ given by
\be
f_{x,\bk} = -4t_1 c_{x/2}c_{y/2}, \quad \bn_{\bk} = (2\lambda s_y, -2\lambda s_x,N_z). \label{eq:buckled-square}
\ee
Here $t_1$ is the nearest neighbor hopping amplitude (i.e., hopping between layers), and we have defined $c_{i/2} \equiv \cos (k_i/2)$ and $s_{i} \equiv \sin k_i$.

In the nonmagnetic state, i.e., when N\'eel order is absent and $N_z=0$, the nonsymmorphic nature of the space group gives rise to symmetry-enforced fourfold degenerate linear crossings at the high symmetry points on the Brillouin zone (BZ) boundary~\cite{Young:2015p126803}. Since these band crossings are described by a gapless Dirac Hamiltonian, which directly follows from expanding the lattice Hamiltonian around the high symmetry points (see Appendix~\ref{app:dirac}), the buckled square lattice of Fig.~\ref{fig1}(b) realizes a symmetry-enforced Dirac semimetal phase. Antiferromagnetic order with N\'eel vector along $\hat z$ gaps the Dirac points on the BZ boundary and gives rise to an antiferromagnetic insulator at half filling~\cite{Wang:2017p115138,Smejkal:2017p106402}, with energy bands still twofold degenerate due to $PT$ symmetry. Applying a displacement field $D_z$ splits the bands and removes this degeneracy, as shown in Fig.~\ref{fig1}(c), and leads to a nonzero orbital magnetization $M_z$ given by Eq.~\eqref{eq:M_z}.

\begin{figure}
	\includegraphics[width=\columnwidth]{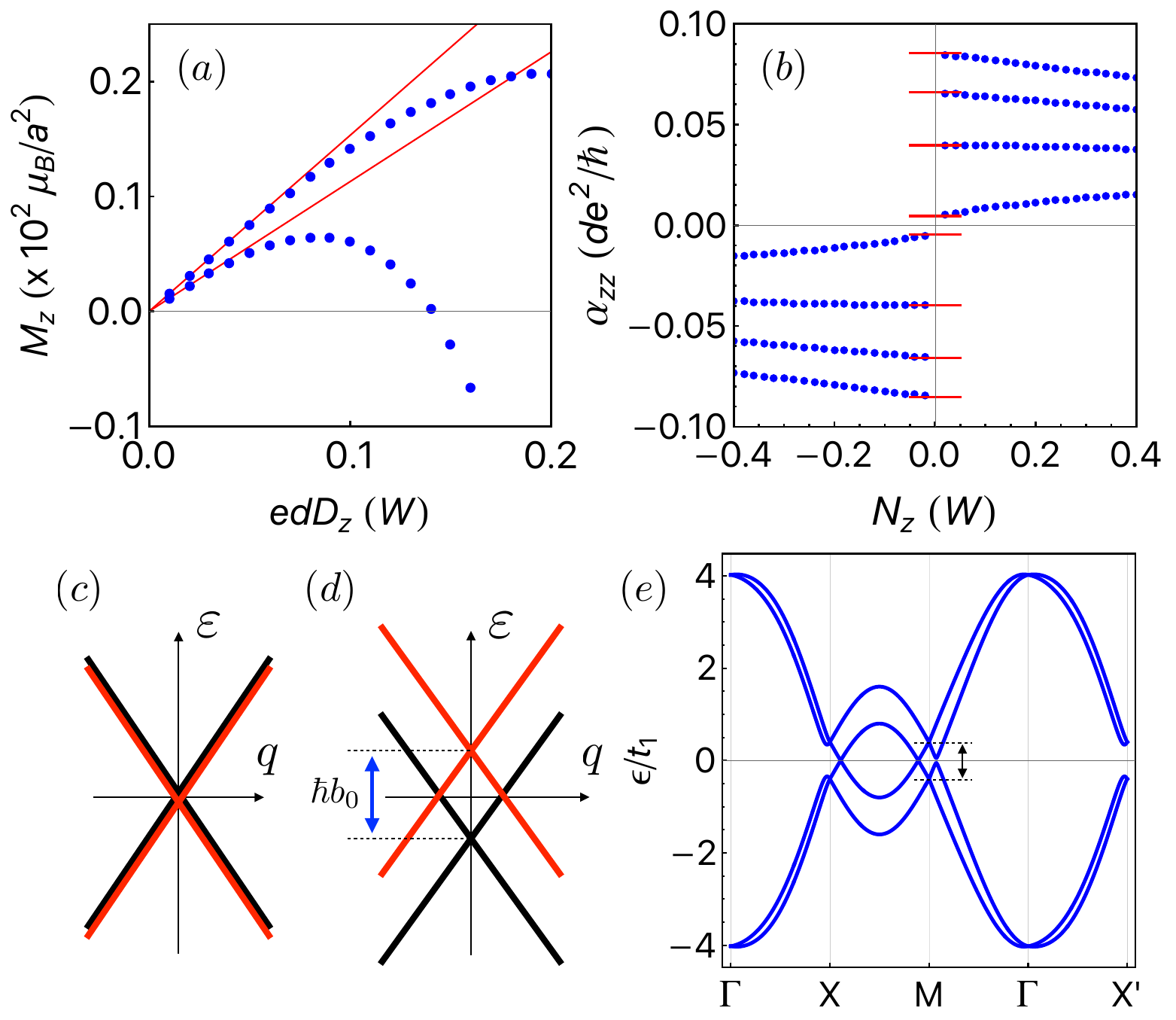}
	\caption{(a) Orbital magnetization as a function of $D_z$, calculated directly from Eq.~\eqref{eq:M_z} (blue dots) and from a linear approximation using Eq.~\eqref{eq:alpha_zz} (red lines). The two sets of curves correspond to $N_z/W=0.2,0.4$. Note that the shown values of $M_z$ have been enlarged by a factor $10^2$.  (b) Polarizability $\alpha_{zz}$ [Eq.~\eqref{eq:alpha_zz}] as a function of $N_z$. Different curves correspond to $\lambda/W=0.6,0.8,1.0,1.2$ (bottom to top for $N_z>0$; top to bottom for $N_z<0$). (c) Sketch of a fourfold Dirac fermion in two dimensions. (d) A perpendicular displacement field $D_z$ shifts the twofold Dirac points in energy by an amount $\hbar b_0 = edD_z$. (e) Energy bands of the buckled square lattice model in the presence of a displacement field $\Delta=edD_z  = 0.2W$ but with $N_z=0$. The resulting energy shift of the Dirac crossings at $M$ is indicated. }
	\label{fig2}
\end{figure}

Panel (a) of Fig.~\ref{fig2} shows the orbital magnetization $M_z$ as a function of $D_z$. All energies are measured in units of $W\equiv  \hbar^2/2m_ea^2$, where $a$ is the lattice constant, and we set $t_1/W=1$. As expected, $M_z=0$ when $D_z=0$,  and $M_z$ has a linear onset given by the polarizability of Eq.~\eqref{eq:alpha_zz}. The dotted blue curves are calculated using Eq.~\eqref{eq:M_z} and the red solid curves are linear approximations based on $M_z = \alpha_{zz} D_z$, confirming that Eq.~\eqref{eq:alpha_zz} correctly captures the linear polarizability $\alpha_{zz}$. The two sets of curves correspond to $N_z = 0.2W$ and $N_z = 0.4W$. We observe that deviations from linearity become strong as $edD_z$ approaches $N_z$, since $edD_z= N_z$ marks a closing of the gap, caused by the touching of the dashed bands in Fig.~\ref{fig1}(c) at $X$ and $M$. 

The magnetoelectric polarizability $\alpha_{zz} $ as a function of N\'eel order parameter $N_z$ is shown in Fig.~\ref{fig2}(b), where different curves correspond to different values of the SOC strength $\lambda$. The dependence of the magnetoelectric polarizability on $N_z$ reveals its most significant feature: a discontinuity at $N_z=0$, indicative of a topological contribution to $\alpha_{zz}$. This topological contribution, and hence the discontinuity of $\alpha_{zz} $, originates from the symmetry-enforced Dirac points present in the nonmagnetic state. A direct way to see this is to approximate the integrand of Eq.~\eqref{eq:alpha_zz} by its contributions from the vicinity of the high-symmetry points $\Gamma$, $X$, and $X'$ (see Appendix~\ref{app:dirac}). The resulting approximate value of $\alpha_{zz}$ is shown in Fig.~\ref{fig2}(b) by red vertical markers and demonstrates that $\alpha_{zz}$ is indeed fully determined by the Dirac point contributions as $N_z \rightarrow 0$. We now examine the topological nature of this contribution in more detail. 

{\it Topological magnetoelectric response.}---As alluded to earlier, the form of Eq.~\eqref{eq:alpha_zz} suggests a connection to the topology of the occupied bands. In fact, we may observe that in cases where $f^2 = 0 $ (see the denominator), the integral corresponds to a topological invariant reflecting the winding of $\bn = \bn_\bk$, and evaluates to an integer multiple of $1/2\pi$. In such cases, the magnetoelectric polarizability is an integer multiple of $e^2 d/2\pi \hbar$. Note that this is not a quantization in units of natural constants, but instead depends on microscopic length scale $d$, i.e., the layer separation. In the buckled square lattice model the condition $f^2 = 0 $ is enforced by symmetry at the high symmetry points on the BZ boundary, and is responsible for the occurrence of Dirac crossings in the nonmagnetic state. This points to Dirac fermions (in two dimensions) as the origin of the topological contribution to magnetoelectric polarizability $\alpha_{zz}$. 

\begin{figure}
	\includegraphics[width=\columnwidth]{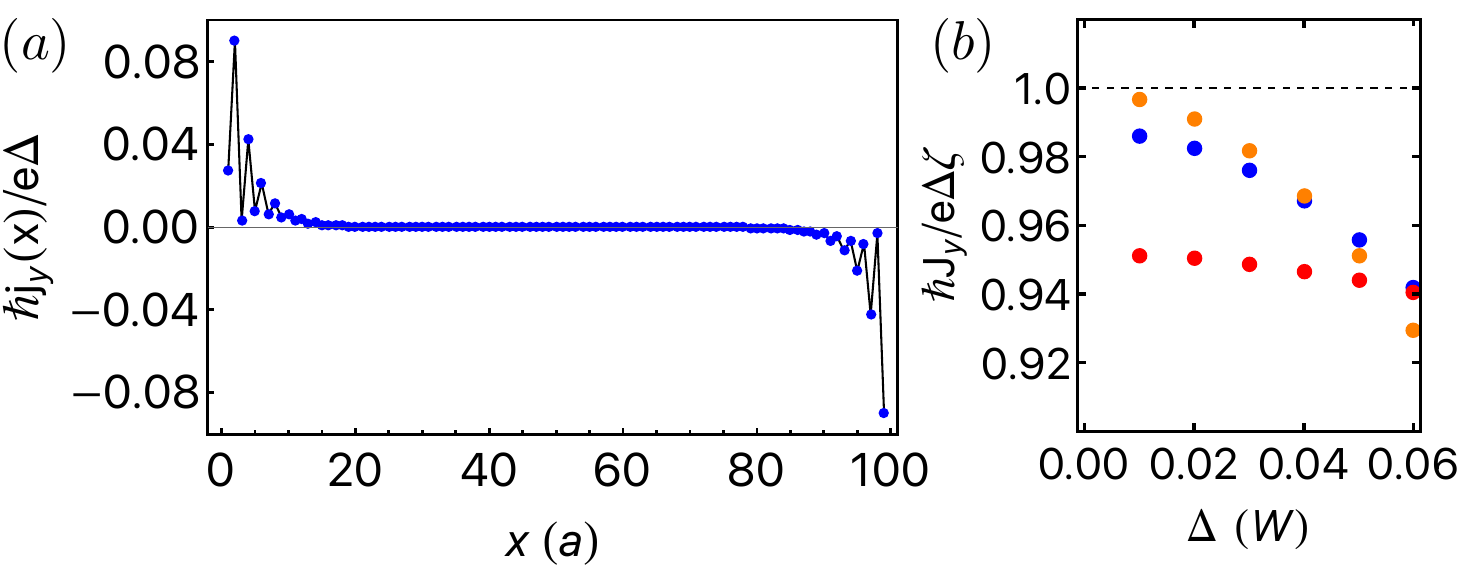}
	\caption{(a) Current density $j_y(x)$ for the buckled square lattice model with open boundary in the $x$ direction ($N_x=100$ sites; $\lambda =0.8W$; $N_z=0.25W$; $\Delta=0.05W$). (b) Integrated current of the right edge as a function of $\Delta$, divided by $\Delta \zeta$ and calculated for $N_z=0.4W$ (red), $N_z=0.2W$ (blue), $N_z=0.15W$ (orange).}
	\label{fig3}
\end{figure}

To establish this connection more firmly, we show how the discontinuity of $\alpha_{zz}$ can be deduced from the general electromagnetic response theory of topological semimetals~\cite{Ramamurthy:2015p085105}. Topological semimetals are gapless electronic phases intermediate between trivial and weak topological insulators~\cite{Armitage:2018p015001,Yan:2017p337,Bernevig:2018p041001,Hirayama:2018p041002,Hasan:2017p1}, and their electromagnetic response theories interpolate between the responses associated with the two distinct insulators~\cite{Ramamurthy:2015p085105}. For Dirac semimetals in two dimensions, the topological response theory predicts an orbital magnetization proportional to separation of the Dirac nodes in energy, which is expressed as 
\be
M_z =-\text{sgn}(m) \frac{e}{2\pi} b_0. \label{eq:Mz_topo}
\ee
Here $b_0$ describes the energy shift in units of frequency, as schematically illustrated in Fig.~\ref{fig3}(c-d), and $m$ is a Dirac mass. (The response is only well defined for massive Dirac theory~\cite{Ramamurthy:2015p085105}.) When applied to a Dirac Hamiltonian of the form $\mathcal H_\bq = \hbar v \tau^z \bq \times \bsigma + N_z \tau_z\sigma^z + \Delta \tau^z $, which describes the buckled square lattice model close to $M=(\pi,\pi)$, then one has $ \Delta = ed D_z = \hbar b_0 $, and therefore $M_z = -\text{sgn}(N_z) (e/2\pi) (ed/\hbar)D_z $. It follows that $\alpha_{zz} = \partial M_z/\partial D_z = - \text{sgn}(N_z)  e^2d/2\pi \hbar $, which constitutes the topological contribution to the magnetoelectric polarizability. 
Its key feature is that it depends only on the sign of the symmetry-breaking mass $N_z$ and does not vanish when ${\mathcal P}$ and ${\mathcal T}$ symmetry are restored as $N_z\rightarrow 0$.  

The general response formula of Eq.~\eqref{eq:Mz_topo} can be applied to all three high symmetry points of the buckled lattice model ($\Gamma$, $X$, and $X'$), as is detailed in Appendix~\ref{app:dirac}. The effect of nonzero $D_z$ on the energy bands on the nonmagnetic state ($N_z=0$) is illustrated in Fig.~\ref{fig2}(e), showing how the Dirac crossings are indeed shifted in energy, as is most clearly visible at $M$. Applying Eq.~\eqref{eq:Mz_topo} to the high symmetry points and summing the resulting contributions yields $\alpha_{zz} =  \text{sgn}(N_z) \zeta e^2d/2\pi \hbar $, with $\zeta \equiv 1-2\lambda/\sqrt{t^2_1+\lambda^2}$, which exactly matches the values shown by red line markers in Fig.~\ref{fig2}(b). This demonstrates that the buckled square lattice provides a minimal model realization of a topological magnetoelectric response associated with the layer degree of freedom.

The topological response of Eq.~\eqref{eq:Mz_topo} can be derived from the effective action~\cite{Ramamurthy:2015p085105}
\be
S[A_i,b_0]  =- \text{sgn}(N_z) \frac{e}{2\pi} \int dt d^2x~\varepsilon^{ij}A_i\partial_j b_0 ,  \label{eq:S_A}
\ee
via variation with respect to the magnetic field. Here $A_i$ is the spatial component of the electromagnetic vector potential. The effective action further describes a current response to a spatially varying $b_0$ given by $j_i  = -\text{sgn}(N_z)(e/2\pi) \varepsilon^{ij}\partial_j b_0 $, a manifestation of the familiar relation $\bj = \boldsymbol{\nabla} \times \bM$. This response implies a current density localized at the edges for a system with open boundaries~\cite{Ramamurthy:2015p085105}. In Fig.~\ref{fig3}(a) we show the current density $j_y(x)$ obtained for the buckled square lattice model, calculated using a cylindrical geometry with open boundary conditions in the $x$ direction, which confirms the existence of edge currents. Integrating the current density on the left edge to obtain the current $J_y = \int_R dx j_y(x)$ should then equal the orbital magnetization $M_z$. In Fig.~\ref{fig3}(e) we show the integrated current $J_y$ as a function of $\Delta$ for three values of $N_z/t_1 = 0.15,0.2,0.4$ (orange, blue, red). Dividing the integrated current by $e\Delta \zeta/\hbar $, we expect that it approaches unity as $\Delta$ and $N_z$ approach zero, since it should equal the topological polarizability $\alpha_zz$ in this limit. This is indeed what Fig.~\ref{fig3}(b) shows \footnote{Note that for smaller values of $N_z$ (which enters as the mass of the Dirac theory and hence sets the localization length of the edge currents) one is limited by finite-size effects of the open boundary system.}. 

\begin{figure}
	\includegraphics[width=\columnwidth]{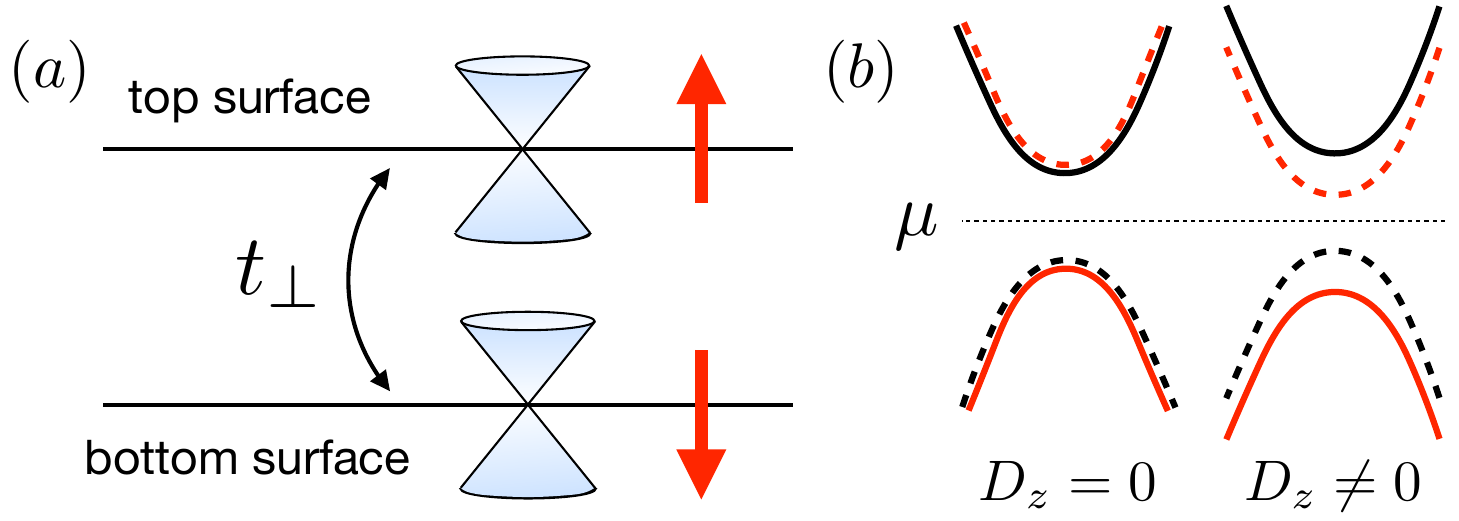}
	\caption{(a) Surface state model of a 2D (thin film) antiferromagnetic TI. Top and bottom surfaces are coupled to surface-normal ferromagnetic order. (b) Schematic energy spectrum of surface state with (left) and without (right) displacement field $D_z$. The chemical potential $\mu$ lies in the gap.}
	\label{fig4}
\end{figure}

{\it Application to magnetic topological insulator.}---As a second example of a bilayer antiferromagnet, consider a 2D antiferromagnetic topological insulator (TI). Thin films of MnBi$_2$Te$_4$ with an even number of septuple layers are realizations of this phase~\cite{Otrokov:2017p025082,Otrokov:2019p416,Rienks:2019p423,Li:2019peaaw5685}. The essential electronic structure of the 2D antiferromagnetic TI can be described by a simple surface state model, shown in Fig.~\ref{fig4}(a), which considers the top and bottom surfaces of the film as the constituents of the bilayer. The Dirac fermions of each surface are coupled to ferromagnetic order with surface-normal orientation, resulting in an overall antiferromagnetic state. Tunneling between the surfaces is captured by a tunneling amplitude $t_\perp$, thus giving rise to the surface model Hamiltonian
\be
H_\bk = \hbar v \tau^z (k_x \sigma^y-k_y \sigma^x ) + t_\perp \tau^x + N_z \tau^z \sigma^z, \label{eq:H-TI}
\ee
which is of the general form \eqref{eq:H_k} with $f_x = t_\perp$ and $\bn = (-  \hbar v k_y,  \hbar vk_x, N_z )$. The energy spectrum, both with and without displacement field $D_z$, is sketched in Fig.~\ref{fig4}(b).

Using Eq.~\eqref{eq:alpha_zz}, it is straightforward to obtain a simple analytical expression for the polarizability $\alpha_{zz}$ of this TI surface state model. We find $\alpha_{zz} = e^2dN_z/2\pi\hbar \sqrt{ t^2_\perp+N^2_z}$, which, in the limit where $N_z \ll t_\perp$, reduces to $\alpha_{zz} \simeq e^2dN_z/2\pi\hbar |t_\perp|$. The magnetoelectric polarizability is thus proportional to $N_z$, as expected based on symmetry arguments, and does not have a topological contribution. The latter is a consequence of $t_\perp$, the tunneling between the surfaces. It is worth noting in passing that the buckled lattice model discussed above can be viewed as a specific 2D lattice regularization of the surface state model, but with key additional symmetry-enforced features on the BZ boundary due to the space group embedding (see Appendix~\ref{app:dirac}).

The 2D magnetic topological insulator has been connected to a number of intriguing topological or magnetoelectric phenomena~\cite{Otrokov:2019p416,Zhang:2019p206401,Liu:2020p522,Lee:2019p012011,Deng:2020p895,Deng:2021p36,Gao:2021p521}, in particular also the layer Hall effect~\cite{Gao:2021p521}. The latter is observed in the presence of a perpendicular displacement field when the chemical potential is in the band, rather than in the gap as in Fig.~\ref{fig4}(b). The magnetoelectric effect proposed here is thus an electric field-induced response specific to the insulating regime. 

{\it Discussion and conclusion.}---We conclude by summarizing the main results, discussing a number of implications, and mentioning connections to materials. Our principal result is the description of a magnetoelectric response associated with the layer pseudospin degree of freedom in 2D ${\mathcal P}$${\mathcal T}$-symmetric antiferromagnets. For a family of microscopic models we have determined the orbital magnetization in response to a perpendicular replacement, obtained the corresponding magnetoelectric polarizability, and demonstrated that the latter can have a topological contribution originating from Dirac fermions. 

The discussion of the magnetoelectric polarizability has been entirely framed in terms of the magnetization response to a perpendicular displacement field, given by $\alpha_{zz} = \partial M_z / \partial D_z$. This has enabled the connection with topological response theory of Dirac semimetals in 2D. A Maxwell relation relates $\partial M_z / \partial D_z$ to the electric pseudospin polarization response $P_z$ to a magnetic field $B_z$, where the former is given by $P_z =ed \langle  \tau^z \rangle$ in the model defined by Eq.~\eqref{eq:H_k}. Therefore, all results referring to $\alpha_{zz}$ equally apply to $ \partial P_z / \partial B_z$. 

As far as connections to materials are concerned, we note that the tetragonal magnet CuMnAs can be understood as a stack of buckled square lattices~\cite{Smejkal:2017p106402,Wang:2017p115138}. A unit layer of magnetic Mn sites thus realizes a buckled square lattice model with the correct space group. Magneto-crystalline anisotropy leads to easy-plane N\'eel order, however, such that moments lie in the plane. 

{\it Acknowledgements.}---We gratefully acknowledge insightful discussions with Daniel Agterberg, Cristian Batista, Charles Kane, Rafael Fernandes, Jeroen van den Brink, and David Vanderbilt. H.R. and J.W.F.V. were supported by the U.S. Department of Energy under Award No. DE-SC0025632. C.O. acknowledges partial support by the Italian Ministry of Foreign Affairs and International Cooperation PGR12351 (ULTRAQMAT) and from PNRR MUR Project No. PE0000023-NQSTI (TOPQIN).

\appendix

\section{Orbital magnetization of four-band model \label{app:Mz-4-band}}

In this Appendix we provide additional details of the derivation of Eqs.~\eqref{eq:M_z} and \eqref{eq:alpha_zz} in the main text. 

As a first step, we rewrite the general expression for the orbital magnetization in terms of band energies $\varepsilon_{ n}$ and projectors $P_n$ onto the corresponding eigenspaces. This yields
\begin{align}
M_z  & = \frac{e  \epsilon_{ij}}{2\hbar}\int \frac{d^2\bk}{(2\pi)^2} \sum_n \text{Im} \langle \partial_i u_{ n} | H_\bk + \varepsilon_{\bk n} | \partial_j u_{ n} \rangle, \\
& \equiv \frac{e}{\hbar} \int \frac{d^2\bk}{(2\pi)^2} \; \mathcal I_\bk.
\end{align}
with the integrand $\mathcal I_\bk$ given by 
\be
\mathcal I_\bk = \sum_{n,m} \frac{\varepsilon_{n} + \varepsilon_{m}}{(\varepsilon_{ n}-\varepsilon_{m})^2} \text{Im}\text{Tr}[P_n  (\partial_xH) P_m (\partial_yH)] .
\ee
Here $\sum_n$ is a sum over occupied bands and $\sum_m$ is a sum over \emph{unoccupied} bands. In this form $\mathcal I_\bk$ is manifestly expressed in terms of projectors and derivatives of the Hamiltonian $\partial_i H$. Recall that the Hamiltonian is parametrized as (momentum dependence suppressed)
\be
H_\bk = f_{x} \tau^x+ f_{y} \tau^y+  \tau^z \bn \cdot \bsigma +\Delta \tau^z, \label{app:H_k}
\ee
and the energies may be determined in a simple way by first evaluating $H^2_\bk$. This gives
\be
H^2_\bk = (f^2 + \bn^2 + \Delta^2) +2\Delta \bn \cdot \bsigma,
\ee
from which it follows that the four energy bands are given by 
\begin{align}
\varepsilon_1 &= - \varepsilon_4 = -\sqrt{f^2 +(|\bn|+ \Delta)^2}  = - \varepsilon_+ \label{app:E1}\\
\varepsilon_2 &= -\varepsilon_3= -\sqrt{f^2 +(|\bn|- \Delta)^2}  = - \varepsilon_- \label{app:E2} 
\end{align}
with $\varepsilon_\pm$ defined as $\varepsilon_\pm  \equiv\sqrt{f^2 +(|\bn|\pm \Delta)^2}$. As indicated in the main text, the ordering of energy bands assumes $\Delta >0$. 

Computing $M_z$ also requires the projectors onto the eigenspaces. These can be found using the method outlined in Ref.~\onlinecite{Graf:2021p085114}. To apply this method, we write the Hamiltonian as $H = \bh \cdot \boldsymbol{\Lambda}$, where $\boldsymbol{\Lambda}$ is a vector which collects the fifteen generators of the SU(4) Lie algebra. We choose a representation of these generators given by
\be
\Lambda_{\alpha} = (\btau, \tau^x \bsigma, \tau^y \bsigma, \tau^z \bsigma ,\bsigma )/\sqrt{2}.
\ee
The generators satisfy the commutation and anti-commutation relations
\begin{align}
[\Lambda_\alpha  ,\Lambda_\beta] & = 2i f_{\alpha\beta\gamma} \Lambda_\gamma, \\
 \{ \Lambda_\alpha  ,\Lambda_\beta \} &= \delta_{\alpha\beta} \mathbb{1}+ 2d_{\alpha\beta\gamma} \Lambda_\gamma,
\end{align}
where $f_{\alpha\beta\gamma}$ and $d_{\alpha\beta\gamma}$ are the antisymmetric and symmetric structure constants. In our particular case, the only nonzero components of $\bh$ are
\begin{align}
(h_1,h_2,h_3) &= \sqrt{2}(f_x,f_y,\Delta) \\
 ( h_{10},h_{11},h_{12}) &= \sqrt{2}(n_x,n_y,n_z).
\end{align}
The projectors onto the four eigensubspaces take the general form $P_n  =(\mathbb{1} +  \bb_n \cdot \boldsymbol{\Lambda})/4$, where $n$ labels the four bands [see Eqs.~\eqref{app:E1} and \eqref{app:E2}]. In our particular problem the vectors $\bb_n$ are given by
\be
\bb_n = \frac{\bh}{ \varepsilon_n} +\frac{\bh_{\star}}{ \varepsilon^2_n-\bh^2/2}+\frac{\bh_{\star\star}}{\varepsilon_n( \varepsilon^2_n-\bh^2/2)},
\ee
where $\bh_{\star}$ is defined as 
\be
\bh_{\star} \equiv \bh \star \bh , \qquad (\bh \star \bh)^\alpha =  h^\beta h^\gamma d_{\alpha\beta\gamma},
\ee
and $\bh_{\star\star}$ is defined as $\bh_{\star\star}\equiv ( \bh \star \bh) \star \bh $. (Note that the $\bk$ dependence is suppressed here.) 

The derivatives of the Hamiltonian are simply derivatives of $\bh$, i.e., $\partial_i H = \partial_i \bh \cdot \boldsymbol{\Lambda}$. The integrand of the orbital magnetization $\mathcal I_\bk$ then becomes
\begin{multline}
\mathcal I_\bk =  \frac{\varepsilon_{1} + \varepsilon_{3}}{(\varepsilon_{ 1}-\varepsilon_{3})^2} \text{Im}\text{Tr}[P_1  (\partial_xH) P_3 (\partial_yH)] \\
  + \frac{\varepsilon_{2} + \varepsilon_{4}}{(\varepsilon_{ 2}-\varepsilon_{4})^2} \text{Im}\text{Tr}[P_2  (\partial_xH) P_4 (\partial_yH)] .
\end{multline}
Using that 
\be
 \frac{\varepsilon_{1} + \varepsilon_{3}}{(\varepsilon_{ 1}-\varepsilon_{3})^2} =  -\frac{\varepsilon_{+} - \varepsilon_{-}}{(\varepsilon_{ +}+\varepsilon_{-})^2}=-  \frac{\varepsilon_{2} + \varepsilon_{4}}{(\varepsilon_{ 2}-\varepsilon_{4})^2},
\ee
the integrand reduces to 
\begin{multline}
\mathcal I_\bk =  -\frac{\varepsilon_{+} - \varepsilon_{-}}{(\varepsilon_{ +}+\varepsilon_{-})^2} \left\{  \text{Im}\text{Tr}[P_1  (\partial_xH) P_3 (\partial_yH)] \right.  \\
\left.  -  \text{Im}\text{Tr}[P_2  (\partial_xH) P_4 (\partial_yH)] \right\}.
\end{multline}
The traces can be evaluated in a straightforward way; collecting the results yields, after simple algebra,
\be
\mathcal I_\bk = 2\Delta \frac{\bn \cdot \partial_x \bn \times \partial_y \bn}{( \varepsilon_++ \varepsilon_-)^3} \times  \frac{4 \Delta^2- (\varepsilon_++ \varepsilon_-)^2 }{ \varepsilon_+ \varepsilon_-},
\ee
which is the result discussed in the main text.

\section{Continuum Dirac model of buckled square lattice \label{app:dirac}}

This Appendix provides details of the low-energy continuum description of the buckled square lattice model discussed in the main text. Such a description is obtained by an expansion of the Hamiltonian near the high-symmetry points where the fourfold Dirac points are located. The three Dirac points are located at $M=(\pi,\pi)$, $X=(\pi,0)$, and $X'=(0,\pi)$. It is sufficient to focus on $M$ and $X$, since $X'$ is related to $X$ by symmetry. 

{\it Expansion around $M$.} We expand the full lattice Hamiltonian $H_\bk$ in small momentum $\bq$ around $M$ and find 
\be
\mathcal H^{M}_\bq =  \hbar  v \tau^z(q_x \sigma^y - q_y \sigma^x )+ N_z \tau^z \sigma^z + \Delta \tau^z ,  \label{app:H^M_q}
\ee
where $v = 2\lambda/\hbar$ is the Dirac velocity. When $\Delta=0$ this Dirac Hamiltonian describes two massive Dirac fermions (with mass $N_z$), one in each $\tau^z=\pm 1$ sector. When $N_z=0$, it describes two massless Dirac fermions shifted in energy by an amount $\Delta$. 

{\it Expansion around $X$.}  When expanded in small momenta $\bq$ with respect to the $X$ point, we find that Hamiltonian reads as
\be
\mathcal H^{X}_\bq =  2t_1 q_x \tau^x  + 2\lambda q_x \tau^z \sigma^y + 2\lambda q_y \tau^z \sigma^x + N_z \tau^z \sigma^z + \Delta \tau^z. \label{eq:H^X_q}
\ee
At the $X$ point the Dirac velocities in the $x$ and $y$ direction are not equal, thus giving rise to an anisotropic Dirac theory, as are given by
\be
v_x = \frac{2}{\hbar}\sqrt{t^2_1 + \lambda^2}, \quad v_y = 2\lambda/\hbar , \label{app:v-X}
\ee
We perform a unitary rotation of the Dirac Hamiltonian defined as 
\be
U = \cos\tfrac{\theta}{2} \mathbb{1}- i \sin\tfrac{\theta}{2} \tau^y \sigma^y = e^{- i \theta \tau^y \sigma^y/2  },
\ee
with $\theta$ defined via the relations 
\be
 \sin\theta = \frac{2t_1}{\hbar v_x}, \; \cos\theta = \frac{2\lambda}{\hbar v_x}, \label{app:theta-X}
\ee
such that the rotated Hamiltonian takes the form
\begin{multline}
U^\dagger \mathcal H^{X}_\bq U =  \tau^z (v_xq_x \sigma^y + v_y q_y  \sigma^x) +N_z \tau^z \sigma^z \\
+ \Delta(  \cos \theta \tau^z-\sin \theta \tau^x\sigma^y).  \label{app:H^X_q-rotate}
\end{multline}
From this form it follows that when $\Delta=0$, the Dirac Hamiltonian at $X$ again describes two massive Dirac fermions (with mass $N_z$), one in each $\tau^z=\pm 1$ sector. Note that there is a meaningful difference with Eq.~\eqref{app:H^M_q}, however, since the sign of the $q_y\sigma^x$ term is flipped. This implies that, relative to Eq.~\eqref{app:H^M_q}, in each $\tau^z=\pm 1$ sector the sign of the Berry curvature of the valence and conduction band is reversed. 

When $N_z=0$ but $\Delta \neq 0 $, Eq.~\eqref{app:H^X_q-rotate} shows there are two distinct terms with different effect on the Dirac points.  The term of interest here is $\Delta \cos \theta \tau^z$, which represents a shift in energy of the two Dirac fermions. Since $\cos \theta  = \lambda / \sqrt{t^2_1 + \lambda^2}$, the shift in energy at $X$ depends on $\lambda$, the strength of spin-orbit coupling.

{\it Magnetization and polarizability from the topological response theory.} We are now in a position to determine the orbital magnetization and the corresponding polarizability using the general topological response formula of Eq.~\eqref{eq:Mz_topo}. The latter equation applies to all three high symmetry points and total orbital magnetization is the sum of all three contributions, which is given by
\be
M_z = -\text{sgn}(N_z) \frac{e }{2 \pi} ( \chi^{\phantom{\dagger}}_M b^M_0+\chi^{\phantom{\dagger}}_{X} b^X_0+\chi^{\phantom{\dagger}}_{X'} b^{X'}_0 ). 
\ee
Here $\chi= \pm 1$ is the chirality associated with the $\tau^z=+1$ Dirac fermion flavor, which is inferred from \eqref{app:H^M_q} and \eqref{app:H^X_q-rotate}. One finds $ \chi^{\phantom{\dagger}}_M =1$ and $ \chi^{\phantom{\dagger}}_X  =  \chi^{\phantom{\dagger}}_{X'} = -1$. Furthermore, we find that
\be
b^M_0 = \frac{\Delta}{\hbar }, \quad b^X_0=b^X_0 = \frac{ \Delta \cos \theta}{\hbar },
\ee
such that 
\begin{align}
M_z &= -\text{sgn}(N_z) \frac{e }{2 \pi} \frac{\Delta(1-2\cos\theta)}{\hbar} \\
&\equiv -  \text{sgn}(N_z) \frac{e }{2 \pi} \frac{\zeta \Delta }{\hbar}.
\end{align}
Here $\zeta$ captures the way in which the three high symmetry points contribute. It depends on the angle $\theta$ defined in Eq.~\eqref{app:theta-X} and reads explicitly as
\be
\zeta =  1- \frac{2\lambda}{\sqrt{t^2_1 + \lambda^2}}. \label{app:zeta}
\ee
Note that for the values of $\lambda$ used in Fig.~\ref{fig2} $\zeta$ is negative, such that $M_z$ is positive. 

{\it Nontopological contribution: Expansion around $\Gamma$.}  It is also useful to perform a long-wavelength expansion around the zone center $\Gamma$. Here we do not expect a topological contribution to $\alpha_{zz}$. Expanding the full lattice Hamiltonian $H_\bk$ with $\Delta =0$ in small momentum $\bq$ around $\Gamma$ and find 
\be
\mathcal H^{\Gamma}_\bq =   -\hbar v \tau^z(q_x \sigma^y - q_y \sigma^x ) -4t_1\tau^x +  N_z \tau^z \sigma^z  ,  \label{app:H^G_q}
\ee
where $v = 2\lambda/\hbar$ is the Dirac velocity (equal to the Dirac velocity at $M$). This Dirac Hamiltonian describes two massive Dirac fermions,  with compatible masses $-4t_1$ and $N_z$. Importantly, the theory is gapped even when $N_z=0$, which is due to the nearest neighbor hopping $t_1$. Applying Eq.~\eqref{eq:alpha_zz} to compute $\alpha_{zz}$, one finds the contribution
\be
\alpha^\Gamma_{zz} = \frac{e^2d }{2\pi\hbar} \frac{N_z}{\sqrt{ 16t^2_1+N^2_z}}, \label{app:alpha_zz-G}
\ee
which captures the behavior of $\alpha_{zz}$ as a function of $N_z$ shown in Fig.~\ref{fig2}(d) in the regime where $\lambda \gtrsim t_1$ (i.e., the regime in which the continuum Dirac model is quantitatively appropriate).

{\it Connection to magnetic TI model.} The continuum expansion of the buckled square lattice model at $\Gamma$, as given by Eq.~\eqref{app:H^G_q}, can be compared to the magnetic TI surface state model of Eq.~\eqref{eq:H-TI}. These Hamiltonians are in fact equal, and we can see that the nearest neighbor hopping $t_1$ plays the role of $t_\perp$ in the surface state model. As a result, the nontopological contribution to the magnetoelectric polarizability given by \eqref{app:alpha_zz-G} is indeed equal to the polarizability obtained for the TI surface state model (after replacing $4t_1$ with $t_\perp$). Since the continuum expansion of the buckled square lattice model at $\Gamma$ reproduces Eq.~\eqref{eq:H-TI}, we can think of the former as a lattice regularization of the TI surface state model. This lattice regularization is special, however, since it has symmetry-enforced Dirac crossings on the Brillouin zone boundary (in the nonmagnetic state), which are a result of space group symmetries and thus of the specific space group embedding. As detailed in the main text, it is these symmetry-enforced features at the high symmetry points on the Brillouin zone boundary which are responsible for the topological contribution to the magnetoelectric polarizability.

\end{document}